\documentclass[11pt]{article}
\usepackage{amsfonts,amsmath,amssymb,bm,hyperref,graphicx}
\makeatletter
\newcommand\figcaption{\def\@captype{figure}\caption}
\newcommand\tabcaption{\def\@captype{table}\caption}

\makeatother \textheight 24 true cm \textwidth 16.5 true cm
\title{Dark matter pair production at the ILC in the littlest Higgs model with T-parity}
\author{{Qing-Peng Qiao$^{1,2}$\footnote{E-mail: xxqqp@mail.nankai.edu.cn},
Bin Xu$^3$} \\
{\small \it 1. Department of Physics, Henan Normal University,
453003, Xinxiang, China} \\{\small \it 2. Department of Physics,
Henan Institute of Education, 450046, Zhengzhou, China} \\{\small
\it 3.  Department of Mathematics and Information Sciences, North
China Institute}\\{\small \it of Water Conservancy and Hydroelectric
Power, 450011, Zhengzhou, China}}
\begin{document}
\maketitle {\bf Abstract}

In the Littlest Higgs model with T-parity, the heavy photon
($A_{H}$) is supposed to be a possible dark matter (DM) candidate. A
direct proof of validity of the model is to produce the heavy photon
at accelerator. In this paper, we study the production rate of $ e^
+ e^ - \to A_{H}A_{H}$ at the international $e^ + e^ -$ linear
collider (ILC) in the Littlest Higgs model with T-parity and show
the distributions of the transverse momenta of $A_{H}$. The
numerical results indicate that the heavy photon production rate
could reach the level of $10^{-1} fb$ at some parameter space, so it
can be a good chance to observe the heavy photon via the pair
production process with the high luminosity at the ILC (500
$fb^{-1}$) . We know that DM is composed of weakly-interacting
massive particles (WIMPs), so the interactions with standard model
(SM) particles are weakness, how to detect heavy photon at a
collider and distinguish it from other DM candidates are simply
discussed in the final sector of the paper.

{\bf PACS numbers:} 13.66.Fg, 13.66.Hk, 95.35.+d

{\bf Keywords:} Littlest Higgs, T-parity, dark matter, ILC

\newpage

\section{Introduction}

~~~~SM of particle physics, including the strong and electroweak
interactions, is the $SU(3)_C \otimes SU(2)_L\otimes U(1)_Y $ gauge
model which has been extensively tested during the past 30 years.
The great success of SM in so-far all the fields of particle physics
manifests that it is obviously a valid theory. However, on other
aspect, because so many parameters in the theory are free and need
to be phenomenologically determined, one can believe it to be only
an effective theory. Besides all the questions which would promote
theorists to look for a more fundamental theory, there exists a
famous problem, namely the hierarchy. This is because the Higgs of
SM gets a mass through a ``bare" mass term and interactions with
other particles, but because of its scalar nature, the mass it gets
through interactions is as large as the largest mass scale in the
theory. In order to get the relative low Higgs mass, the bare mass
must be fine-tuned to several tens of an expressive places so as to
accurately cancel the very big interaction terms. Meanwhile, the
astrophysical observations, for example, the rotation curves of
galaxies \cite{Begeman:1991iy} and the gravitational lensing effects
\cite{Massey:2007wb} show that about $23\%$ of the energy density of
the universe is composed of DM. One of the most fundamental problems
in cosmology and particle physics today is what the nature of DM is.
A flood of discussions point out that the DM should have the
following characters: non-luminous, non-baryonic, non-relativistic,
and electrically neutral \cite{Roszkowski:1991ng, Roszkowski:1999ts,
Primack:2001ia, Chen:2003bn, Ma:2004nw}. Extensive astronomical
evidence indicates that DM is a stable heavy particle that interacts
with SM particles weakly. That is to say, such a particle is a WIMP
\cite{Bertone:2004pz}. The microscopic composition of DM remains a
mystery and SM does not offer an appropriate candidate to account
for DM, however, many theories which extend SM contain new particles
with the proper properties to play the role of DM.

As mentioned above, the new physical models are needed to slove the
fine-tuning problem and provide the candidates of DM.

The best known example that can solve the above-mentioned problems
is a supersymmetry (SUSY) model with R-parity \cite{Dreiner:2005rd}.
The minimal supersymmetric standard model (MSSM) with R-parity
\cite{Goldberg:1983nd, Ellis:1983ew} is such a model. Supersymmetric
partners are introduced, the contribution of each supersymmetric
partner cancels off the contribution of each ordinary particle, so
the fine-tuning doesn't exist anymore. On the other hand, a discrete
gauge symmetry named R-parity is included, so the lightest
neutralino, a Majorana fermion ($\tilde \chi _1^0$) is supposed to
be the lightest stable SUSY particle and stand as the DM component,
where the neutralinos ($\tilde \chi _1^0,{\rm{ }}\tilde \chi
_2^0,{\rm{ }}\tilde \chi _3^0,{\rm{ }}\tilde \chi _4^0$) are the
lightest particles in the SUSY models, they are the fermionic SUSY
partners of the neutral gauge and CP-even Higgs bosons.

A little Higgs model \cite{Cheng:2004yc,Low:2004xc} with T-parity
 \cite{Skiba:2003yf, Cheng:2003ju} provides a successful
alternative to the SUSY with R-parity in solving the problems of the
SM mentioned above. In the initial little Higgs model (LHM), Higgs
originates as a pseudo-Nambu-Goldstone boson (PNGB) of a
spontaneously broken global symmetry. The global symmetry is
definitely broken by the interactions of two sets, with each set
preserving an unbroken subset of the global symmetry. The Higgs is
an exact NGB when either set of couplings is no longer present. The
hierarchy problem is explained by adopting new heavy particles at
the TeV scale with the same spins as the corresponding SM particles.
One-loop quadratic divergences of the Higgs boson mass are canceled
by these new particles. Unfortunately, these initial models suffered
from strict restrictions from the precision electroweak fits.

An ideal resolution to solve the problem is to apply a $Z_2$
discrete symmetry named T-parity (analogous to the R-parity in
SUSY), by requiring SM particles to be even and the new heavy
particles to be T-parity odd, forbidding all tree-level corrections
to the electroweak precision tests. If the T-parity was
conservation, the lightest T-odd particle (LTP) of a LHM would be
stable and not decay, and it will play the role of the DM candidate.

The littlest Higgs model with T-parity (LHT) \cite{Hubisz:2004ft,
Birkedal:2006fz} is such a type model, it is a modification of the
original Littlest Higgs Model. One of the unique signatures of the
theory is the existence of a neutral minimum weight heavy $U(1)$
gauge boson: heavy photon $A_H$ and some other special particles.
Experimentally searching them would provide direct evidence for
judging the validity of the theory. There are several ways to study
DM, for example, one can use the relativistic mean field theory to
determine the nuclear form factor for the detection of dark matter
\cite{Chen:2011xp}. Astronomical observations and restrictions may
provide another way. In our previous work, we have studied the heavy
photon time-evolution in the LHT \cite{Qiao:2007rc}. Finally, There
is a good opportunity to combine theoretical models with the
upcoming DM direct and indirect detection experiments, results from
the present and future colliders in the TeV range. Because all the
new particles are very heavy, they escaped the detection (if exist)
in previous accelerators. By a general analysis, their masses, just
like the particles predicted by other models, may be in TeV regions,
so that one can expect to observe them at the high energy colliders.
The production of heavy photon pair at the large hadron collider
(LHC), $pp \to {A_H}{A_H} + X$ and the associated production of
$Q_HA_H$ at the LHC have been discussed in Refs.
\cite{Hubisz:2004ft} and \cite{Chen:2006ie} respectively, where
$Q_H$ is the partner of the light quark. Because of the complex
background, the LHC might be difficult to confirm a theory, so the
ILC will be a proper tool to better understand the properties of the
LHT. In this work, we analyze the production of the heavy photon
pair at the ILC.

The paper is organized as follows. The numerical results of the
production rate and the distributions of the transverse momenta are
presented in Sec. II where all input model parameters are explicitly
listed. Our discussions on various aspects and conclusions are made
in the last section.

\section{Theoretical formulation and numerical results}

~~~~First, let us concisely recall the relevant characteristics of
the LHT. The LHT is based on a non-linear $\sigma$-model describing
a global $SU(5)/SO(5)$ symmetry breaking, which takes place at an
energy scale $\Lambda\sim4\pi f\sim$10 TeV. The SM Higgs doublet is
generally considered to be a subset of the Goldstone bosons
associated with the breaking. The symmetry breaking also breaks the
assumed embedded local gauge symmetry $[SU(2)\times U(1)]^2$
subgroup down to the diagonal subgroup  $SU(2)_L\times U(1)_Y$,
which is identified as the ordinary SM electroweak gauge group. The
additional gauge structure leads to four extra gauge bosons at the
TeV scale, $W_H^ \pm$, $Z_H$ and $A_H$. The diagonal Goldstone
bosons of the matrix $\Pi$ are ``eaten" to become the longitudinal
degrees of freedom of the heavy gauge bosons with odd T-parity at
the scale $f$, the masses of them are
\begin{eqnarray}
M(Z_H) \approx M(W_H^\pm) \approx gf, ~{\rm{ }}M({A_H}) \approx
\frac{{g'}}{{\sqrt 5 }} \approx 0.16f,
\end{eqnarray}
where $g$ and $g'$ are the gauge couplings of SM $SU(2)_L$ and
$U(1)_Y$. After electroweak symmetry breaking at the scale $v \ll
f$, the masses of the new heavy gauge bosons as well as the SM gauge
boson masses receive corrections of order $v^2/f^2$ and could be
written as
\begin{flalign}
 \begin{split}
M(Z_H)=M(W_H^\pm)=gf(1 - \frac{{{v^2}}}{{8{f^2}}}) \approx 0.65f,\\
~{\rm{ }}M({A_H})= \frac{{fg'}}{{\sqrt 5 }}(1 -
\frac{{5{v^2}}}{{8{f^2}}}) \approx 0.16f,
 \end{split}
 \end{flalign}
since the masses of the other T-odd particles are generically of
level $f$, the $A_H$ can be assumed to be the LTP and be regarded as
an ideal candidate of the WIMP cold DM.

The mirror fermions, acquire masses through a Yukawa-type
interaction
\begin{eqnarray}
\kappa f({\bar \Psi _2}\xi \Psi ' + {\bar \Psi _1}{\Sigma _0}\Omega
{\xi ^\dag }\Omega \Psi ')
\end{eqnarray}
whereas $\Psi _1$, $\Psi _2$ are the fermion doublets and $\Psi '$
is a doublet under $SU(2)_2$.

One fermion doublet ${\Psi _H} = \frac{1}{{\sqrt 2 }}({\Psi _1} +
{\Psi _2})$ acquires a mass $\kappa f$, which is a free parameter,
with the natural scale set by $f$. Specifically, the T-odd heavy
partners of the SM leptons acquire the following masses $\sqrt{2}
\kappa_l f$ \cite{Cacciapaglia:2009cv}, where the $\kappa_l$ is the
flavor independent Yukawa coupling. The T-odd fermion mass for both
lepton and quark partners will be assumed to exceed 300 GeV to avoid
the colored T-odd particles from being detected in the squark
searches at the Tevatron.

In the LHT model, the coupling term in the Lagrangian related to the
heavy photon of our work is written as \cite{Hubisz:2004ft,
Birkedal:2006fz}:
\begin{eqnarray}
A_H^\mu {\tilde L_i}{L_j}: ~i\frac{e}{{10{C_W}{S_W}}}({S_W} -
5{C_W}{(\frac{v}{f})^2}{x_h}){\gamma ^\mu }{P_L}\delta_{ij},\
\end{eqnarray}
where $\tilde L$ is the heavy lepton of the LHT and $L$ is the SM
lepton, and $e$ is the electromagnetic coupling constant, ${x_h} =
\frac{5}{4}\frac{{gg'}}{{5{g^2} - {{g'}^2}}}$, $v =
\frac{{2{M_W}{S_W}}}{e}$, $f=1000$, ${\rm{ }}{P_L} = \frac{
\displaystyle {1 - {\gamma_5}}}{\displaystyle 2}$. We identify $g$
and $g'$ with the SM $SU(2)$ and $U(1)_Y$ gauge couplings,
respectively. $S_W$ and $C_W$ are sine and cosine of the Weinberg
angle, respectively.

The Feynman diagrams responsible for the process of $e^ + e^ - \to
A_{H}A_{H}$ at the tree-level are shown in Fig.\ref{Fig.1}.
\begin{center}
\includegraphics[width=5.8cm]{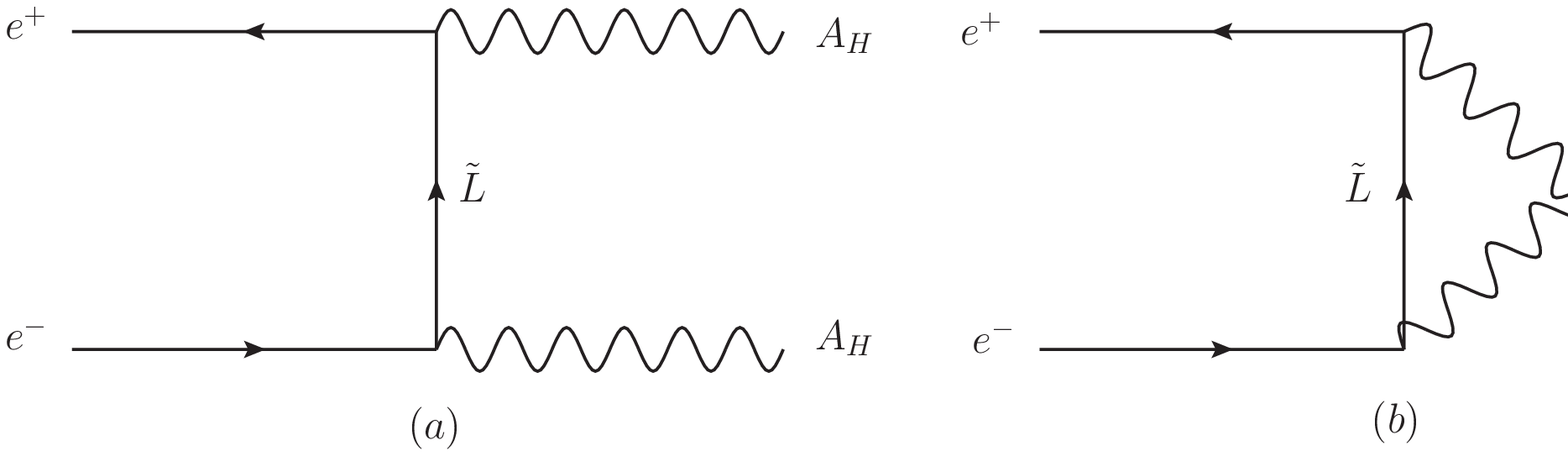}
\figcaption{\label{Fig.1} The Feynman diagrams of the process $e^ +
e^ - \to A_{H}A_{H}$.}
\end{center}

With the Lagrangian, the amplitude of the process can be written
directly:
\begin{flalign}
 \begin{split}
 &M = {M_a} + {M_b}\\
 & =-i(\frac{e}{{10{C_W}{S_W}}}({S_W} -
5{C_W}{(\frac{v}{f})^2}{x_h}))^2\bar v({P_1})(\frac{1}{{
 P_{24}^2 - m_{\tilde{L}}^2}}
{\gamma ^\mu }\\
&~~{\gamma ^\rho}{\gamma^\nu }{P_{24\rho }}+ \frac{1}{{P_{23}^2 -
m_{\tilde{L}}^2}}{
  \gamma ^\nu }{\gamma ^\rho }{\gamma ^\mu }{P_{23\rho }})
{P_L}u({P_2})\epsilon _\mu ^*({P_3})\epsilon _\nu ^*({P_4}),
 \end{split}
 \end{flalign}
where $P_{23}=P_2-P_3$, $P_{24}=P_2-P_4$, an $\epsilon$ is the
polarization vector for the heavy photon gauge boson.

The differential cross section is given by:
\begin{eqnarray}
\frac{{d\sigma }}{{dt}} = \frac{1}{{2!}}\frac{1}{{64\pi
s}}\frac{1}{{{{\left| {{{\vec p}_1}} \right|}^2}}}\bar \Sigma
{\left| M \right|^2},
\end{eqnarray}
where s and t are the Mandelstam variables, $dt = 2\left| {{{\vec
p}_1}} \right|\left| {{{\vec p}_3}} \right|d\cos \theta$.

With the above production amplitudes and differential cross section,
we can directly obtain the production cross section of the process.
For the numerical computations of the cross section, we need the
electroweak fine-structure constant $\alpha$ to satisfy
$\alpha(M_Z)=\frac{e^2}{4\pi}=\frac{1}{128}$
\cite{Nakamura:2010zzi}.

There are three free parameters involved in the production
amplitudes: the heavy photon mass $m_{AH}$, the mass of heavy lepton
$m_{\tilde{L}}$ and the energy of the center-of-mass frame
$\sqrt{s}$. Concretely, in order to expose possible dependence of
the cross section on these parameters, we take two groups of values:
$\sqrt{s}=$500, 1000 GeV, $m_{\tilde{L}}=$300, 500, 700 GeV
respectively, the $m_{AH}$ takes 100 to 250 GeV in Fig.\ref{Fig.2}
and 100 to 300 GeV in Fig.\ref{Fig.3}. The numerical results of the
cross sections are plotted in Fig.\ref{Fig.2} and Fig.\ref{Fig.3}.
\begin{center}
\includegraphics[width=8cm]{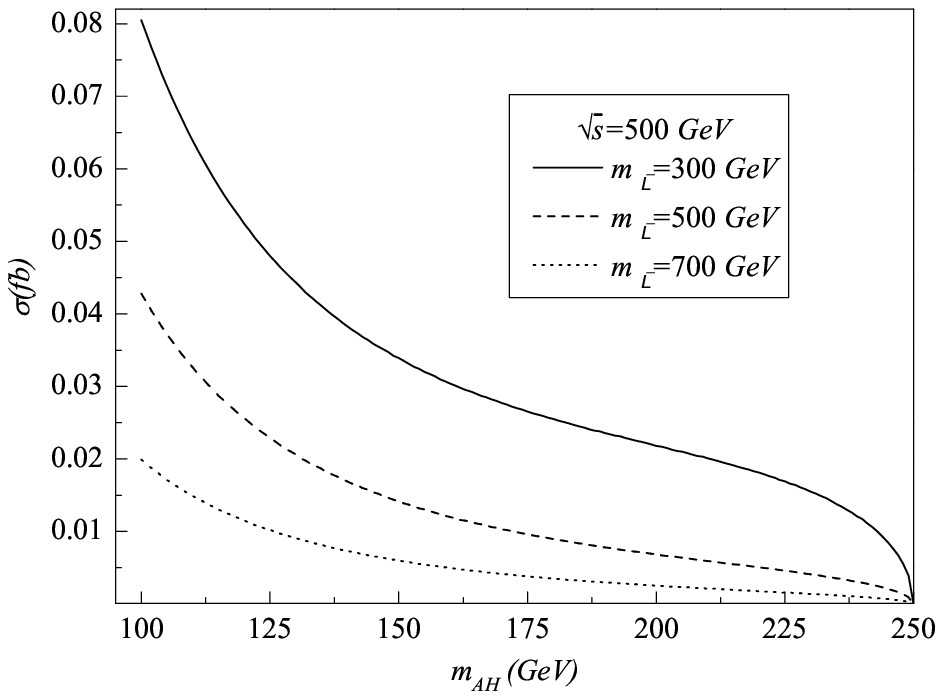}
\figcaption{\label{Fig.2} The dependence of the cross section of $
e^ + e^ - \to A_{H}A_{H}$ on heavy photon mass $m_{AH}$ (100$\sim
$250 GeV) for $\sqrt{s}$=500 GeV and $m_{\tilde{L}}=$300, 500, 700
GeV at the ILC.}
\end{center}
\begin{center}
\includegraphics[width=8cm]{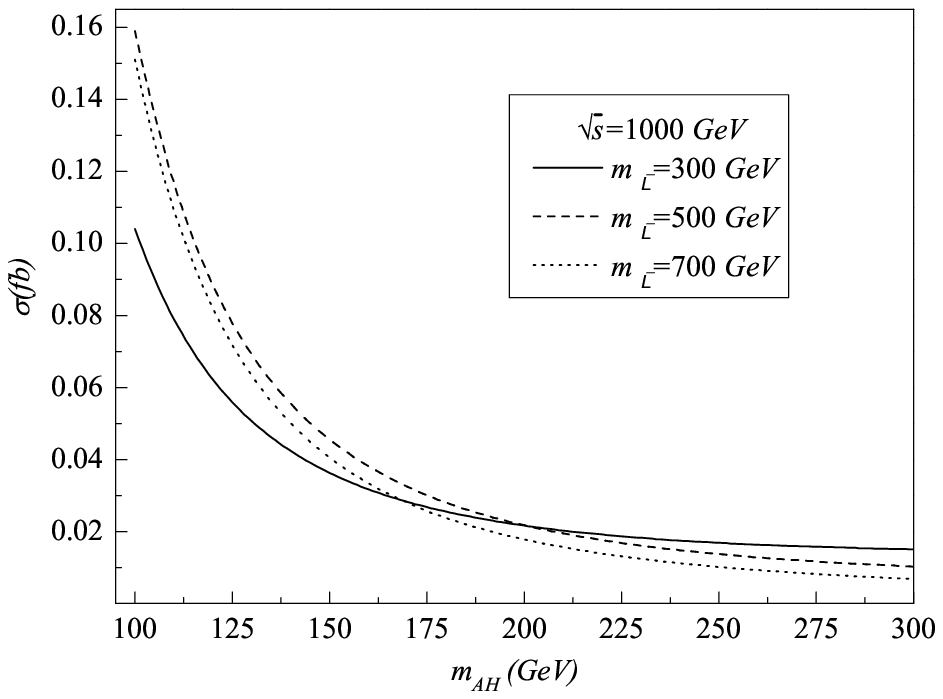}
\figcaption{\label{Fig.3} The dependence of the cross section of $
e^ + e^ - \to A_{H}A_{H}$ on heavy photon mass $m_{AH} $ (100$\sim
$300 GeV) for $\sqrt{s}$=1000 GeV and $m_{\tilde{L}}$=300, 500, 700
GeV at the ILC.}
\end{center}
\begin{center}
\includegraphics[width=8cm]{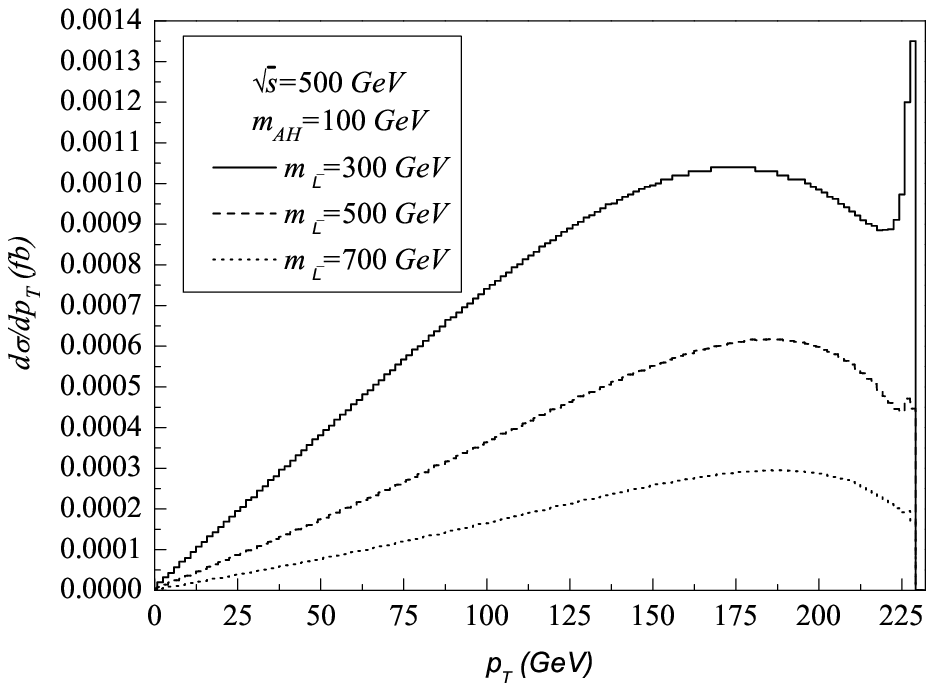}
\figcaption{\label{Fig.4} The distributions of the transverse
momenta of final state ($p_{T}$) for the process $ e^ + e^ - \to
A_{H}A_{H}$ with $\sqrt{S}$=500 GeV.}
\end{center}
\begin{center}
\includegraphics[width=8cm]{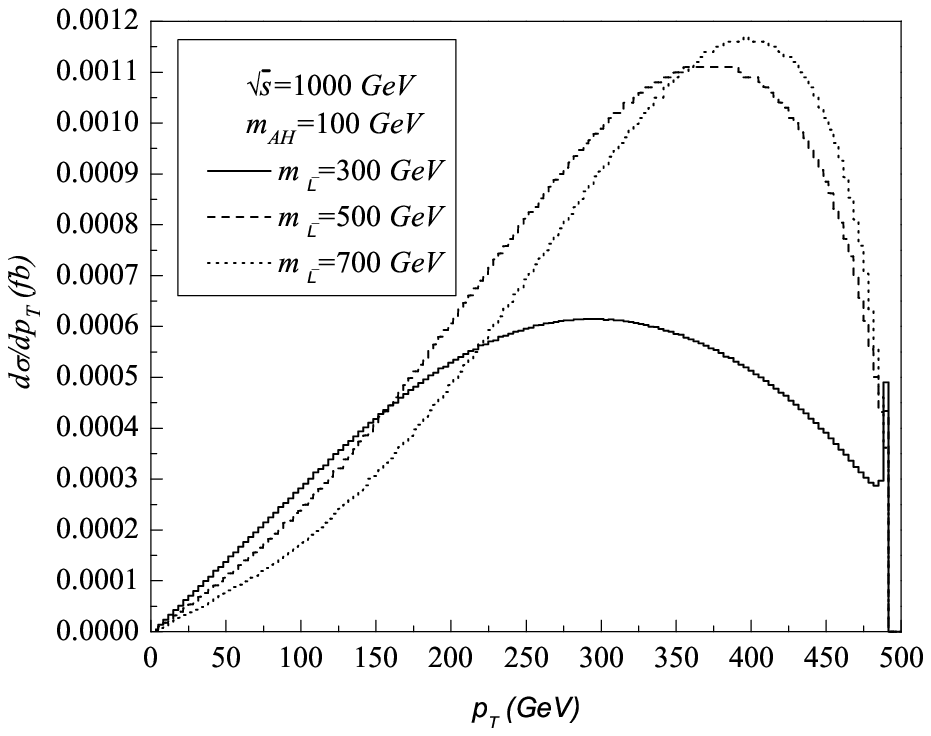}
\figcaption{\label{Fig.5} The distributions of the transverse
momenta of final state ($p_{T}$) for the process $ e^ + e^ - \to
A_{H}A_{H}$ with $\sqrt{S}$=1000 GeV.}
\end{center}

From these figures, we observe that the production rate decreases
sharply as $m_{AH}$ increases due to the constraint from the phase
space of the final state involving the heavy photon pair. The
dependence of the cross section on $\sqrt{s}$ in the range of
parameters that we use is apparent: when $\sqrt{s}$ becomes large
the cross section increases obviously. The relations between the
cross section and $m_{\tilde{L}}$, are the other way around, when
$\sqrt{s}$ is fixed as 500 GeV, $m_{\tilde{L}}$ becomes large, the
cross section decreases, but when $\sqrt{s}$ takes the value of 1000
GeV, the situations are more complicated, one could see the concrete
relations from the figures.

In Fig.\ref{Fig.4} and Fig.\ref{Fig.5}, we present the transverse
momentum distributions of heavy photon for $\sqrt{S}$=500 and 1000
GeV with $m_{AH}$ equal to 100 GeV, respectively. From the figures,
we could see that the differential cross section increases with the
transverse momentum until the value reaches the maximum at the point
that $p_T$ equals a certain value, then it begins to slide.

\section{Discussions and conclusions}

~~~~The advantages of observing the process of $ e^ + e^ - \to
A_{H}A_{H}$ at the ILC are obvious. First, astrophysical probes
provide a way to study the characteristics of DM, however,
astrophysical observations are unable to determine and examine the
exact properties of a DM particle and it mostly remains for collider
physics to reveal the nature of DM. Second, the LHC and the ILC may
well turn out to be the DM factories and the ILC background is
relatively clean, so this paper is a proper supplement to the Refs.
\cite{Hubisz:2004ft,Chen:2006ie}. Finally, the heavy photon is the
LTP and must be produced in pairs, constraint from the final state
phase space of the above process is alleviated compared with other
processes.

The combination of cosmology and high energy collider seems to be a
good idea \cite{Khlopov:1998ux, Belotsky:1998pv}, however, at the
same time, we have to solve some problems.

The first problem is  how to distinguish the $A_H$ pair of the LHT
from other DM pair production at the ILC, for example, the lightest
neutralino ($\tilde \chi _1^0$) of SUSY with R-parity
\cite{Goldberg:1983nd, Ellis:1983ew}. Since different DM particles
have different reaction channels and probabilities to be detected by
the detectors of the ILC, which offer a way to distinguish the $A_H$
from others. In the MSSM with R-parity, tree-level production
cross-sections of ${e^ + }{e^ - }\to \tilde \chi _i^0  \tilde \chi
_j^0$ have been studied systematically, readers who are interested
in these processes and want to know more about them can see the
paper \cite{Bartl:1986hp, Carena:2006gb}, so we don't discuss the
topic in detail here.

There is a more serious question we have to face, unlikely the
production of heavy photons in $e^+ e^-$ annihilation of
six-dimensional SUSY QED \cite{Fayet:1985kt}, DM does not carry
either color or electric charge, the direct detection of DM at a
cllider is very hard. In the collider experiments, DM would be like
neutrinos and therein they would escape the detector without
depositing energy in the experimental devices, causing an obvious
imbalance of momentum and energy in collider events. DM would
manifest itself as missing energy, however, DM isn't the only source
responsible for the missing energy. Limited calorimeter resolution,
uninstrumented regions of the detector, and additional energy of
cosmic rays must all be considered thoroughly in a collider
experiment. All of these factors strongly complicate the
investigation of DM in the collider experiments. For conservation of
momentum, the sum of all momenta transverse to the beam direction
must equal zero. Thus, we can ascertain the missing energy by
measuring the energy deposited in each calorimeter cell of a
detector. If all of the above uncertainties and the background
caused by SM neutrinos were been are subtracted and the vector sum
of all the transverse momenta is not equal to zero, we could claim
that something invisible is produced, the undetected particle(s) may
be the DM candidate(s) (such as a heavy photon).

There is a better way to observe the heavy photon pair production.
They can  be observed via radiative production $ e^ + e^ - \to
A_{H}A_{H}\gamma$. In this process, the signal is a single high
energetic photon and missing energy, carried by the heavy photons.
It would be our next work \cite{Qiao:2011sw}, we'll discuss the
problem later in detail.

Although it is very hard to detect the DM, $ e^ + e^ - \to
A_{H}A_{H}$ is the leading order process of the DM production in the
Littlest Higgs model with T-parity at the ILC, so we need to predict
the production rate. As a conclusion, we find that the production
rate of $ e^ + e^ - \to A_{H}A_{H}$ could reach the level of
$10^{-2} fb$ at some small mass parameter space, which is a bit more
larger than the process of $e^ + e^ - \to A_{H}Z_{H}$
\cite{Asakawa:2009qb}, the heavy gauge bosons $A_H$ and $Z_H$ are
produced with the cross section of 1.9 $fb$ at the center of mass
energy of 500 GeV and the large mass of heavy $Z_H$ boson(369 GeV)
suppresses the final state phase space. The production of $A_{H}$
may be observable as the missing energy at the ILC thinks to its
high energy and luminosity. However, for some uncertain reasons, the
missing energy related to the $A_{H}$ is very difficult to be
accurately determined, so identification of the DM production at the
ILC is very hard. We, so far, are not equipped with such expertise
and ability to handle the complex analysis, but will definitely
cooperate with our experimental colleagues and experts in this field
to carry out a detailed analysis.

\section*{Acknowledgments}

~~~~This work is supported by the National Natural Science
Foundation of China (No. 11075045) and the Natural Science
Foundation of Education Department of Henan Province (No.
2011A140005).

\end{document}